\documentclass[sigconf, anonymous=false, screen=true]{acmart}

\renewcommand\footnotetextcopyrightpermission[1]{}
\usepackage[T1]{fontenc}

\usepackage{booktabs}
\usepackage{pgfplots}
\pgfplotsset{compat=newest}
\usepackage{xcolor}
\usepackage{siunitx}
\usepackage{multirow}
\usepackage{xspace}
\usepackage{balance}
\usepackage[font=rm]{caption}
\usepackage{verbatim}
\usepackage{graphicx}
\usepackage{tabularx}
\usepackage{enumitem}
\usepackage{shortvrb}
\usepackage{tikz}
\usepackage{subfig}
\usepackage[boldmath]{numprint}

\usepackage[noend]{algpseudocode}
\usepackage[linesnumbered
			,vlined]{algorithm2e}

\makeatletter
\newcommand{\removelatexerror}{\let\@latex@error\@gobble}
\makeatother

\setlist[itemize,1]{leftmargin=3mm,itemsep=1mm}
\setlist[enumerate,1]{leftmargin=5mm,itemsep=1mm}

\definecolor{DarkGray}{gray}{.45}

\newcommand{\parag}[1]{\vspace{0.15cm}\noindent\textbf{\textsf{#1.}}}

\newcommand{\mytablescale}{0.9}
\newcommand{\myscreenshotscale}{0.4}
\newcommand{\mycaption}[1]{\caption{{\rm{#1}}}}

\SetKwIF{If}{ElseIf}{Else}{{\textsf{if}}}{}{\textsf{else if}}{{\textsf{else :}}}{}
\SetKwFor{For}{\textsf{for}}{}{}
\SetKwFor{While}{\textsf{while}}{}{}
\newcommand{\code}[1]{\textsf{\textbf{#1}}}
\newcommand{\func}[1]{\textsf{{#1}}}

\usepackage{xcolor, soul}

\usepackage{pbox}
\usepackage{makecell}
\usepackage{tikz}
\usetikzlibrary{calc}

\newcommand{\aol}{\textsf{AOL}}
\newcommand{\msn}{\textsf{MSN}}
\newcommand{\ebay}{\textsf{EBAY}}

\newcommand{\pref}{prefix-search}
\newcommand{\conj}{conjunctive-search}

\newcommand{\one}{\textsf{Fwd}}
\newcommand{\two}{\textsf{FC}}
\newcommand{\three}{\textsf{Heap}}
\newcommand{\four}{\textsf{Hyb}}
\newcommand{\comptrie}{\textsf{Trie}}
\newcommand{\compfc}{\textsf{FC}}

\newcommand{\RNum}[1]{\uppercase\expandafter{\romannumeral #1\relax}}

\newcommand{\seq}{\mathcal{S}}

\newcommand{\extract}{\textup{\textsf{Extract}}}
\newcommand{\locate}{\textup{\textsf{Locate}}}
\newcommand{\locateprefix}{\textup{\textsf{LocatePrefix}}}
\newcommand{\access}{\textup{\textsf{Access}}}
\newcommand{\rmq}{\textup{\textsf{RMQ}}}

\newcommand{\Nextgeq}{\textup{\textsf{NextGeq}}}

\newcommand{\method}[1]{{\sf{#1}}}

\newcommand{\var}[1]{\mbox{\emph{#1}}}

\newcommand{\vb}{{\method{VB}}}

\newcommand{\bic}{{\method{BIC}}}
\newcommand{\ef}{{\method{EF}}}
\newcommand{\pef}{{\method{PEF}}}

\newcommand{\dint}{\method{DINT}}

\newcommand{\optvb}{{\method{OptVB}}}

\newcommand{\simples}{{\method{Simple16}}}

\begin{document}

\fancyhead{}

\title{Efficient and Effective Query Auto-Completion}

\author{Simon Gog}
\affiliation{%
  \institution{eBay Inc.}
}
\email{sgog@ebay.com}

\author{Giulio Ermanno Pibiri}
\affiliation{%
  \institution{ISTI-CNR} 
}
\email{giulio.ermanno.pibiri@isti.cnr.it}

\author{Rossano Venturini}
\affiliation{%
  \institution{University of Pisa} 
}
\email{rossano.venturini@unipi.it}

\begin{abstract}
Query Auto-Completion (QAC) is an ubiquitous feature
of modern textual search systems, suggesting possible
ways of completing the query being typed by the user.
Efficiency is crucial to make the system have
a real-time responsiveness when operating in the
million-scale search space.
Prior work has extensively advocated the use of
a trie data structure for fast prefix-search operations
in compact space.
However, searching by prefix has little discovery power in that
only completions that are prefixed by the query
are returned. This may impact negatively the effectiveness
of the QAC system, with a consequent
monetary loss for real applications like Web Search Engines
and eCommerce.

In this work we describe the implementation that empowers a new
QAC system at eBay, and discuss its efficiency/effectiveness
in relation to other approaches at the state-of-the-art.
The solution is based on the combination of
an inverted index with succinct data structures,
a much less explored direction in the literature.
This system is replacing the previous implementation based on
Apache SOLR that was not always able to meet the
required service-level-agreement.


\end{abstract}

\maketitle

\section{Introduction}\label{sec:introduction}

The Query Auto-Completion (QAC) problem we consider
can be formulated as follows.
Given a collection $\seq$ of scored strings
and a \emph{partially-completed} user query $Q$, find the
top-$k$ scored completions that match $Q$ in $\seq$.
For our purposes, a completion is a \emph{full} (i.e., completed)
query for which the search
engine, that indexes a large document collection, returns a
relevant and non-empty recall set.
The collection $\seq$ is usually a query log consisting in
several million user queries seen in the past, with
scores taken as a function of the frequencies of the queries.
The straightforward approach of suggesting the ``most popular queries''~\cite{bar2011context}
(i.e., the ones appearing more often)
works well for real-world applications like eBay search.

At eBay, the QAC system helps users to formulate queries to
explore 1.4 billion live listings \emph{better} (e.g., with less spell errors) and \emph{faster} (as we also include relevant category constraints).
This is important for desktop users but, in particular, for the 
growing number of users of mobile devices.
In fact, as QAC is expected to happen instantaneously, these systems
have a low-millisecond service-level-agreement (SLA).
The previous system implemented at eBay, based on Apache SOLR\footnote{\url{https://lucene.apache.org/solr}},
was not always able to meet the SLA and had a sub-optimal
memory footprint.
This motivated the development of eBay's new QAC system.


In this paper, we share the basic building blocks of the retrieval
part of the system. In short, it is based on a combination of 
succinct data structures, an inverted index, and tailored retrieval
algorithms.
We also provide
an open-source implementation in C++ of the presented techniques --
available at \url{https://github.com/jermp/autocomplete} --
with a reproducible experimental setup
that includes state-of-the-art baselines.

Lastly, we remark that a production version of this system\footnote{This system includes spell correction and business logic.
Both parts add latency but also were improved by the presented techniques.}
was implemented in eBay's Cassini search framework
and can serve about 135,000 query per seconds at 50\% CPU
utilization on a 80-core machine.
(The 99-quantile latency is below 2 milliseconds and 
the average latency is about 190 $\mu$s.)

\section{Related work}\label{sec:related}

The QAC problem has been studied rather extensively
since its popularization by Google around 2004.
The interested reader can refer to the general
surveys by~\citet{cai2016survey} and~\citet{krishnan2017taxonomy}
for an introduction to the problem.

Following the taxonomy given by~\citet{krishnan2017taxonomy},
we have two major auto-completion query
modes -- \emph{prefix-search} and \emph{multi-term prefix-search} --
that have been implemented and are in wide-spread use.
Our own implementation at eBay is no exception, thus these
are the query modes we also focus on.

Informally speaking,
searching by prefix means returning strings
from $\seq$ that are prefixed by the concatenation of the
terms of $Q$;
a multi-term prefix-search identifies
completions where all the terms of $Q$
appear as prefixes of some of the terms of the completions,
regardless their order.
We will describe and compare these two
query modes in details in Section~\ref{sec:solutions}.
Prefix-search is supported efficiently by representing $\seq$
with a {trie}~\cite{fredkin1960trie}
and many papers discuss this approach~\cite{hsu2013space,bar2011context,mitra2015query,mitra2014user,shokouhi2012time,shokouhi2013learning}.
Multi-term prefix-search is, instead, accomplished via
an inverted index built from the completions in $\seq$~\cite{bast2006type,ji2009efficient}.
In particular, if we assign integer identifiers (docids) to the
completions,
an inverted list is materialized for each term that
appears in $\seq$ and stores the identifiers of the completions
that contain the term.
(As we are going to illustrate in the subsequent sections,
\emph{how docids are assigned to the completions}
is fundamental
for the efficiency of the inverted index and, hence, of the
overall QAC system.)

For example, if $Q$ is \textsf{``shrimp dip rec''}, then
a plausible completion found by prefix-search could be
\textsf{``\textbf{shrimp dip rec}ipes''}.
A multi-term prefix-search
could return, instead, \textsf{``\textbf{shrimp} bienville \textbf{dip rec}ipe''} or \textsf{``\textbf{rec}ipe for appetizer \textbf{shrimp} chipolte \textbf{dip}''}.
Note that all terms of $Q$ are prefixed
by some terms of these two example completions
but in no specific order.

The focus of this paper is on the query modes,
rather than on the ranking of results.
As already stated,
we consider the popular strategy of ranking the results
by their frequency within a query log.
There have been some studies comparing different
ranking mechanism for a \emph{single} query mode, e.g.,
prefix-search~\cite{di2015comparing}.
However,
little attention was given to the efficiency/effectiveness
trade-off between \emph{different} query modes,
with an exception in this regard being the experimentation
by~\citet{krishnan2017taxonomy}.
They also report significant variations in effectiveness
by varying query mode.

We now briefly summarize two results
that are closely related to the contents of this paper
because both use an inverted index.
\citet{bast2006type} merge the inverted lists into blocks
and store their unions to reduce the number of lists.
As we will see, this is crucial
to sensibly boost the responsiveness of the
QAC system in the case of single-term queries.
For queries involving several terms,
\citet{ji2009efficient} propose an efficient algorithm
to quickly check whether a completion belongs to the union
of a set $L$ of inverted lists.
Instead of trivially computing the union,
the idea is to check whether the terms in the completion
overlap with those corresponding to the inverted lists in $L$.

\parag{Other Approaches}
Although inherently different from the direction we pursue here,
other approaches may include \emph{sub-string search},
where each term of $Q$ can occur as a sub-string of a completion.
However, to the best of our knowledge, there are no
implementations of this query mode but only a discussion by~\citet*{chaudhuri2009extending}.
Also, suggesting $n$-grams from the indexed
documents was found to be very effective
in absence of a query log~\cite{bhatia2011query}.


\begin{figure}[t]
\removelatexerror
\centering

\subfloat[] {
    \scalebox{\mytablescale}{\input{prefix_search.tex}}
}

\subfloat[] {
    \scalebox{\mytablescale}{\input{conjunctive_search.tex}}
}

\mycaption{Auto-Completion algorithms based on {\pref} (a) and {\conj} (b).
\label{alg:search}}
\vspace{-0.5cm}
\end{figure}

\section{Efficient and Effective Query Auto-Completion}\label{sec:solutions}

Here we describe the QAC algorithm
used at eBay -- in essence, based on the
\emph{multi-term prefix-search} query mode that
we are going to call \emph{\conj} from now on.

We begin with an overview of the different steps
involved in the identification of the top-$k$
completions for a query in Section~\ref{sec:query_processing}.
The aim of such section is to introduce
the data structures and algorithms involved in the search.
From Section~\ref{sec:introduction},
recall that we use $\seq$ to denote
the set of scored strings from which
completions are returned.
As we will see next, we have several data structures
built from $\seq$, such as:
(1) a \emph{dictionary}, storing all its distinct
terms;
(2) a representation of the completions that
allows efficient prefix-search;
(3) an inverted-index.
Section~\ref{sec:data_structures} discusses the
implementation details of these data structures.
Lastly, in Section~\ref{sec:conjunctive}
we explain how to implement
{\conj} efficiently.

\subsection{Query Processing Steps}\label{sec:query_processing}

In this section we detail the processing steps that are executed
to identify the top-$k$ completions for a query.
We are going to illustrate the pseudo code given in Fig.~\ref{alg:search} that shows two different solutions
to the QAC problem,
respectively based on {\pref} (a)
and {\conj} (b).

A detail of crucial importance for the search efficiency
is that we do \emph{not} manipulate scores directly,
rather we assign docids to completions in \emph{decreasing-score order}.
(Ties broken lexicographically.)
This implies that if a completion has a smaller docid than another,
it has a ``better'' score as well.
As we are going to illustrate next,
this docid-assignment strategy
substantially simplifies the implementation of 
\emph{both} {\pref} and {\conj}.

With this initial remark in mind,
we now consider an example of $\seq$ in Table~\ref{tab:completions}.
Suppose $k=3$. If the query $Q$ is \textsf{``bm''}, then
the algorithm in Fig.~\ref{alg:pref} based on {\pref}
would return the completions having docid 1, 2, and 4.
(The other implementation in Fig.~\ref{alg:conj} would
return the same results.)
Now, if $Q$ is \textsf{``sport''}, the algorithm in Fig.~\ref{alg:conj} would find 2, 4, and 6, as the top-3 docids.
(Note that the algorithm in Fig.~\ref{alg:pref} is not able
to answer this query.)

\parag{Parsing}
We consider the query as
composed by a set of terms,
each term being a group of characters separated by white spaces.
The query can possibly end without a white space: in this case the
last query term is considered to be incomplete.
The first processing step involves \emph{parsing} the query, i.e.,
dividing it into two separate parts: a \emph{prefix} and a \emph{suffix}.
The suffix is just the last term (possible, incomplete).
The prefix is made up of all the terms preceding the suffix:
each of them, say $t$, is looked up in a \emph{dictionary}
data structure
by means of the operation called $\locate(t)$ that returns the
lexicographic integer id of $t$.
For example, the id of the term \textsf{``sedan''} is 7,
in the example in Fig.~\ref{tab:inverted_index}.
(If term $t$ does not belong to the dictionary
then an invalid id is returned
to signal this event.)

\begin{table}[t]
\centering
\mycaption{An example set of completions seen as strings
and integer sets in (a).
The integer sets are obtained by replacing the terms
with their ids as given by the dictionary in (b).
\label{tab:example}}
\subfloat[]{
\scalebox{0.8}{\begin{tabular}{c l l}
\toprule
docids & completions & sets \\
\midrule
9 & \textsf{audi}              & $\langle 2 \rangle$ \\
6 & \textsf{audi a3 sport}     & $\langle 2,1,8 \rangle$ \\
3 & \textsf{audi q8 sedan}     & $\langle 2,6,8 \rangle$ \\
8 & \textsf{bmw}               & $\langle 3 \rangle$ \\
5 & \textsf{bmw x1}            & $\langle 3,10 \rangle$ \\
1 & \textsf{bmw i3 sedan}      & $\langle 3,4,7 \rangle$ \\
4 & \textsf{bmw i3 sport}      & $\langle 3,4,8 \rangle$ \\
2 & \textsf{bmw i3 sportback}  & $\langle 3,4,9 \rangle$ \\
7 & \textsf{bmw i8 sport}      & $\langle 3,5,8 \rangle$ \\
\bottomrule
\end{tabular}
}
\label{tab:completions}
}
\subfloat[]{
\scalebox{0.8}{\begin{tabular}{c l l}
\toprule
termids & terms & inverted lists \\
\midrule
1  & \textsf{a3}        & $\langle 6 \rangle$ \\
2  & \textsf{audi}      & $\langle 3,6,9 \rangle$ \\
3  & \textsf{bmw}       & $\langle 1,2,4,5,7,8 \rangle$ \\
4  & \textsf{i3}        & $\langle 1,2,4 \rangle$ \\
5  & \textsf{i8}        & $\langle 7 \rangle$ \\
6  & \textsf{q8}        & $\langle 3 \rangle$ \\
7  & \textsf{sedan}     & $\langle 1,3 \rangle$ \\
8  & \textsf{sport}     & $\langle 4,6,7 \rangle$ \\
9  & \textsf{sportback} & $\langle 2 \rangle$ \\
10 & \textsf{x1}        & $\langle 5 \rangle$ \\
\bottomrule
\end{tabular}
}
\label{tab:inverted_index}
}
\vspace{-0.5cm}
\end{table}

\begin{figure*}[t]

%

\subfloat[{\conj}]{
\includegraphics[scale=\myscreenshotscale]{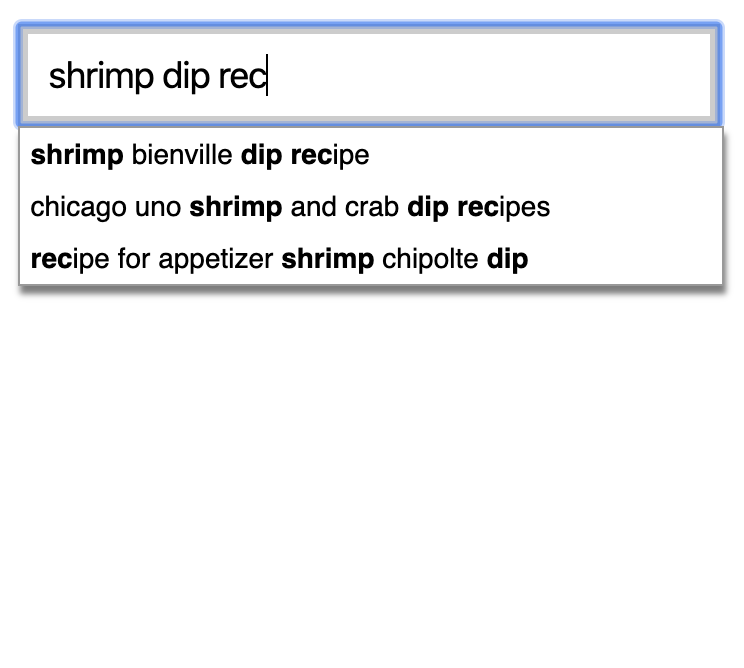}
}
\subfloat[{\pref}]{
\includegraphics[scale=\myscreenshotscale]{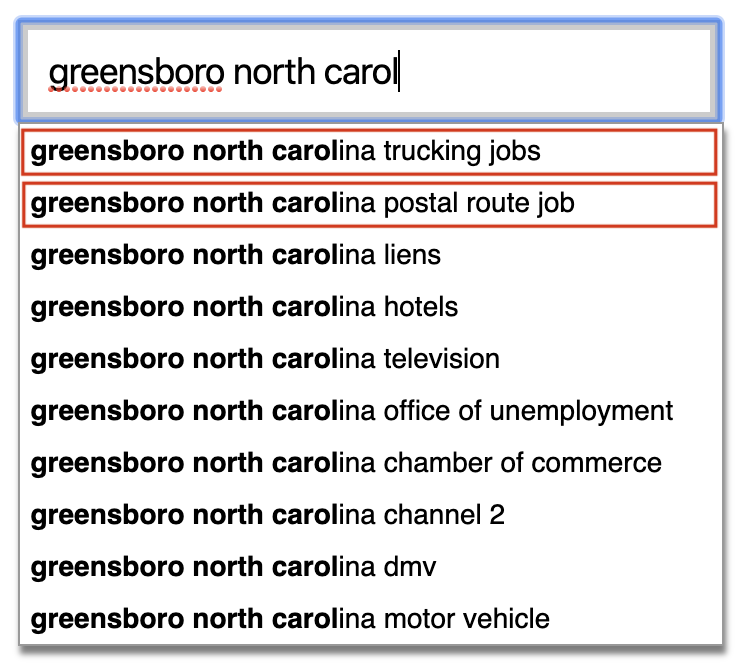}
}
\subfloat[{\conj}]{
\includegraphics[scale=\myscreenshotscale]{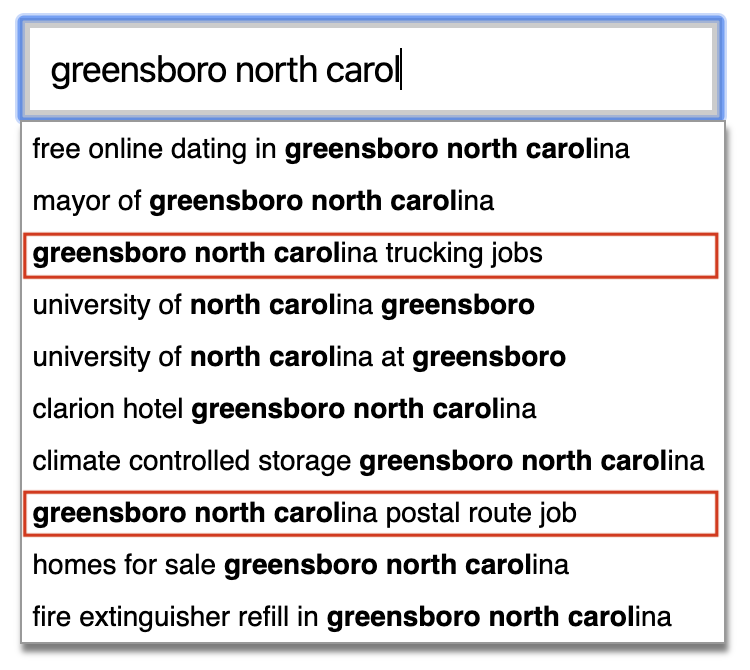}
}

\mycaption{Some example of searches on the {\aol} query log.
In (a), a simple {\pref} returns no results at all,
whereas {\conj} does.
In (b) and (c) we see that both search modes return a full
set of results (top-10), that are significantly different:
only the results enclosed within the boxes are in common.
The example shows that the first two results
returned by {\conj}
have a better score than the first result of {\pref};
the 4th, 5th, 6th, and 7th results of {\conj} have all better
score than the second result of {\pref}. 
\label{fig:screenshots}}
\end{figure*}

\parag{Prefix-search}
Let us assume for the moment
that all terms in the \emph{prefix} are found
in the dictionary.
Let $PS$ indicate the concatenation of \emph{prefix}
and \emph{suffix}.
The \textsf{Complete} algorithm in Fig.~\ref{alg:pref},
based on {\pref},
returns the top-$k$ completions
from $\seq$
that are prefixed by $PS$ and comprises two steps.
First, the dictionary is used to obtain the lexicographic
range $[\ell,r]$ of all the terms that are prefixed by the suffix,
using the operation $\locateprefix(\var{suffix})$.
If there is no string in the dictionary that is prefixed by \var{suffix},
then the range $[\ell,r]$ is invalid and the algorithm returns no results.

Otherwise, the operation $\locateprefix(\var{prefix}, [\ell,r])$ is
executed on a data structure representing the completions of $\seq$.
More precisely, this data structure does not
represent the completions as strings of characters,
but as (multi-) sets of integers,
each integer begin the lexicographic id of a dictionary term.
Refer to Table~\ref{tab:completions} for a pictorial example.
In Fig.~\ref{alg:pref}, we
indicate such data structure with the name \var{completions}.
The operation $\locateprefix(\var{prefix}, [\ell,r])$ returns, instead,
the lexicographic range $[p,q]$ of the completions in $\seq$
that are prefixed by $PS$.
Again, if there are no strings in \var{completions} that
are prefixed by $PS$, then {\pref} fails.

Let us consider a concrete example for the query
\textsf{``bmw i3 s''},
with $\seq$ as in Table~\ref{tab:completions}.
Then $\var{prefix} = \langle 3,4 \rangle$ and
$\var{suffix}$ $=$ ``\textsf{s}''.
The first operation $\locateprefix($``\textsf{s}''$)$
on the dictionary of Table~\ref{tab:inverted_index}
returns $[7,9]$.
The second operation $\locateprefix(\langle 3,4 \rangle, [7,9])$,
instead, returns the range $[6,8]$.

We then proceed with the identification of the top-$k$ completions.
This step retrieves the smallest docids of the completions
whose lexicographic id is in $[p,q]$.
If we materialize the list of the docids
following the lexicographic order of the completions
(such as the column ``docids'' in Table~\ref{tab:completions})
-- say \var{docids} --
then identifying the top-$k$ docids boils down to
support \emph{range-minimum queries} ({\rmq})~\cite{muthukrishnan2002efficient}
over $\var{docids}[p,q]$.
Note that this is possible because of the
used docid-assignment.
We recall that $\func{RMQ}(p,q)$
returns the \emph{position} of the minimum
element in $\var{docids}[p,q]$.
Specifically, we have that $\var{docids}[i] = x$,
where $x$ is the docid
of the $i$-th lexicographically smallest completion.
In other words, \var{docids} is a map
from the lexicographic id of a completion to its docid.
The column ``docids'' in Table~\ref{tab:completions}
shows an example of such sequence.
Continuing the same example as before for the query
\textsf{``bmw i3 s''},
we have got the range $[6,8]$, thus we have to report the $k$
smallest ids in $\var{docids}[6,8]$.
For example, if $k=1$, 
then $\func{RMQ}(6,8) = 6$ and 
we return $\var{docids}[6] = 1$.

\parag{Conjunctive-search}
A {\conj} is a
\emph{multi-term prefix-search} that uses
an inverted index.
At a high-level point of view,
what we would like to do is to identify
completions containing
\emph{all} the terms specified in the \var{prefix}
and \emph{any} term that is prefixed by the \var{suffix}.
We can do this efficiently by computing
the intersection between the
inverted lists of the term ids in the \var{prefix}
and the union of the inverted lists of
all the terms in $[\ell,r]$.
In Section~\ref{sec:conjunctive} we will describe
efficient implementations of this algorithm.

For our example query \textsf{``bmw i3 s''},
the intersection between the inverted lists
in Table~\ref{tab:inverted_index}
of the terms
\textsf{``bmw''} and \textsf{``i3''} gives
the list $X = [1,2,4]$.
Since the range $[\ell,r]$ is $[7,9]$ in our example,
the union of the inverted lists 7, 8, and 9 is $Y = [1,2,3,4,6,7]$.
We then return the docids in $X \cap Y$, i.e., $[1,2,4]$.
In fact,
it is easy to verify that such completions of id 1, 2, and 4,
are the ones having all the two
terms of the prefix and any term among the ones prefixed by \textsf{``s''},
that are \textsf{``sedan''}, \textsf{``sport''} and \textsf{``sportback''}.

Again, observe that since the lists are sorted by docids
\emph{and} we assigned docids in decreasing score order,
\emph{the best results are those appearing before}
as we process the lists from left to right.

We claim that {\conj} is more powerful than {\pref},
for the following reasons.
\begin{itemize}
\item It is not restricted to just the completions that are prefixed by $PS$.
In fact, what if we have a query like \textsf{``i3''} or \textsf{``bmw sport i8''}?
The simple {\pref} is not able to answer.
(No completion
is prefixed by \textsf{``i3''} or \textsf{``bmw sport''}.)
\item There could exist a completion
that is \emph{not} prefixed by $PS$
but has a \emph{better} score than that of some results identified by {\pref}.
Consider the practical examples in Fig.~\ref{fig:screenshots}.
over the {\aol} dataset, one of the publicly
available datasets we use in our experimental analysis.
\item It can also be issued when a query term
is \emph{out} of the vocabulary.
In that case, {\pref} is not able to answer at all
(unless the term is the suffix),
whereas {\conj} can use the other query terms in the prefix.
\end{itemize}

However, as we will experimentally show in Section~\ref{sec:experiments},
this effectiveness does not come for free
in terms of efficiency.

\parag{Reporting}
After the identification of the set of (at most) $k$ top ids,
that we indicate with \var{topk\_ids} in Fig.~\ref{alg:search},
the last step reports the final identified completions,
i.e., those strings having ids in \var{topk\_ids}.
What we need is a data structure supporting the operation
$\access(x)$ that
returns the string in $\seq$ having lexicographic id $x$.
Once we have a completion, that is a (multi-) set of term ids,
we can use the dictionary to {\extract} each term from its id,
hence reconstructing the actual completion's string.

\subsection{Data Structures}\label{sec:data_structures}

In the light of the query processing steps described in
Section~\ref{sec:query_processing} and their
operational requirements,
we now discuss the implementation details
of the data structures they use.

\parag{The Dictionary}
The string dictionary data structure has to support
{\locate}, {\locateprefix} and {\extract}.
An elegant way of representing
in compact space a set of strings while supporting all the
three operations,
is that of using \emph{Front Coding} compression
({\compfc})~\cite{martinez2016practical}.
{\compfc} provides good compression ratios
when the strings share long
common prefixes and remarkably fast decoding speed.

We use a (standard) two-level data structure to represent
the dictionary.
We chose a block size $B$ and compress with {\compfc}
the $\lceil |\var{dictionary}| / (B+1) \rceil$ buckets, with each bucket
comprising $B$ compressed strings (except, possibly, the last).
The first strings of every bucket are stored uncompressed in
a separate \emph{header} stream.
It follows that both operations {\locate} and {\locateprefix}
are supported by binary searching the header strings and then
scanning: one single bucket for {\locate}; or at most two buckets
for {\locateprefix}.
The operation {\extract} is even faster than {\locate} because
only one bucket has to be scanned without any prior binary search.
Clearly the bucket size $B$ controls a space/time trade-off~\cite{martinez2016practical}:
larger values of $B$ favours space effectiveness (less space overhead
for the header), whereas smaller values favours query processing
speed.
In Section~\ref{sec:experiments} we will fix the value of $B$
yielding a good space/time trade-off.

\parag{The Completions}
For representing the completions in $\seq$,
we need a data structure supporting {\locateprefix} which returns
the lexicographic range of a given input string.
(In the following discussion, we assume $\seq$ to be sorted lexicographically.)

As already mentioned in Section~\ref{sec:related},
a classic option is the {trie}~\cite{fredkin1960trie} data structure,
that is a labelled
tree with root-to-leaf paths representing the strings of $\seq$.
In our setting, we need an \emph{integer} trie and we adopt the
data structure described by~\citet*{pibiri2017efficient,pibiri2019handling},
augmented to keep track of the lexicographic range of every node.
More specifically, a node $n$ stores the lexicographic range spanned by
its rooted subtrie: if $\alpha$ is the string spelled-out by the path
from the root to $n$ and $[p,q]$ is the range, then
all the strings prefixed by $\alpha$ span the contiguous range $\seq[p,q]$.
It follows that, for a trie level consisting in $m$ nodes, 
the sequence formed by the juxtaposition of the ranges
$[p_1,q_1] \ldots [p_m,q_m]$
is sorted by the ranges' left extremes, i.e., $p_i < p_{i+1}$
for $i=1,\ldots,m-1$.
Therefore, to allow effective compression is convenient to represent
such sequence as two sorted integer sequences:
the sequence $L$ formed by the left extremes, such that $L[i] = p_i - i$;
and the sequence obtained by considering the range sizes and taking
its prefix sums.
Each level of the trie is, therefore, represented by 4 sorted
integer sequences:
nodes, pointers, left extremes, and range sizes.
Another option to represent $\seq$ is to use {\compfc} compression
as similarly done for
the dictionary data structure.
We now discuss advantages and disadvantages of both options,
and defer the experimental comparison to Section~\ref{sec:experiments}.

\begin{itemize}
\item Tries achieve compact storage because
common prefixes between
the strings are represented \emph{once} by a shared root-to-node path in
the tree.
Prefix coding is clearly better than that of
{\compfc} but the trie needs more redundancy for the encoding
of the tree topology and range information.
\item Although prefix searches in the trie are supported in time linear in
the size of the searched pattern (assuming $O(1)$
time spent per level),
the traversal process is \emph{cache-inefficient}
for long patterns.
The binary search needed to locate the front-coded buckets
is not cache-friendly as well
(and includes string comparisons),
but is compensated by the fast decoding of {\compfc}.
\item Another point of comparison is that of supporting
the $\access(i)$ operation that returns the $i$-th smallest
completion from $\seq$.
This is needed to implement the last step of processing,
that is to report the identified top-$k$ completions as strings.
The trie data structure can not support {\access} without
explicit node-to-parent relationships, whereas {\compfc} offers
a simple solution taking, again, time proportional to $B$.
If we opt to use a trie,
a simple way of supporting {\access} is to explicitly represent the
completions in a \emph{forward} index that is,
essentially, a map from
the docid to the completion.
(The use of a forward index is also crucial for the 
efficient implementation of {\conj}
we will describe in Section~\ref{sec:conjunctive}.)
\end{itemize}

In conclusion, we have two different and efficient ways
of supporting
{\pref} and Reporting:
either a trie plus a forward index,
or {\compfc} compression.


\parag{Range-Minimum Queries}
The identification of the top-$k$ docids in a given lexicographic range
follows a standard approach~\cite{muthukrishnan2002efficient,hsu2013space}.
The algorithm iteratively finds the $k$
smallest elements in $\var{docids}[p,q]$.
To do so, we maintain a min-heap of ranges, each of these keeping
track of the \emph{position} of the minimum element in the range.
At each step of the loop:
(1) we pop from the heap the interval having
the minimum element; (2) add it to the result set;
(3) push onto the heap the two sub-ranges respectively to the
left and to the right on the minimum element.
Correctness is immediate.
To answer a range-minimum query we build and store the \emph{cartesian tree} of
the array \var{docids}. It is well-known that such tree can be represented
in just $2n + o(n)$ bits, with $n$ being the size of the array,
using a succinct encoding such as \emph{balanced parentheses} (BP)~\cite{fischer2011space}.
Since the time complexity of a {\rmq} is $O(1)$ and the heap contains $O(k)$
elements (at each iteration,
we push at most two ranges but always remove one)
it follows that this algorithm
has a worst-case complexity of $\Theta(k \log k)$.

\parag{The Inverted Index}
Inverted indexes are subject of deep study and a wealth of different
techniques can be used to represent them in compressed space~\cite{survey},
while allowing efficient query processing.
What we need is an algorithm for supporting list intersections:
details on how this can be achieved by means of the
Next Greater-than or Equal-to ({\Nextgeq}) primitive are
discussed by many papers~\cite{moffat1996self,survey,ottaviano2014partitioned,slicing}.
The operation $\Nextgeq_t(x)$ returns the element $z \geq x$ from the
inverted list of the term $t$ if such element exists, otherwise
the sentinel $\infty$ (larger than any possible value)
is returned.

\subsection{Multi-term Prefix-search Query Mode: Conjunctive-search}\label{sec:conjunctive}

In this section we discuss efficient implementations
of the {\conj} algorithm introduced in Section~\ref{sec:query_processing}.
We begin our discussion by describing a simple
approach that uses just an inverted index; then highlight
its main efficiency issues and present solutions to
solve them.
Remember that the objective of this query mode is to
return completions that contain \emph{all}
the terms in the \var{prefix}
and \emph{any} term that is prefixed by the \var{suffix}.

\parag{Using an Inverted Index}
A first approach is illustrated in Fig.~\ref{alg:conjunctive_search_heap}.
The idea is to iterate over the elements of the intersection
(lines 7-8)
between the
inverted lists of the \var{prefix} and, for each element,
check whether it appears in \emph{any} of the inverted lists of the terms
in $[\ell,r]$ (inner loop in lines 9-18).
Directly iterating over the elements of the intersection,
rather than computing the \emph{whole} intersection between the
inverted lists,
saves time when the intersection has many results
because we only need the \emph{first},
i.e., smallest, $k$ results.
(Remember that we assign
docids in decreasing score order.)

To implement the check for a given docid \var{x},
we maintain a heap of list iterators:
one iterator for each inverted list.
To be clear, an iterator over a list is an object
that has the capability
of skipping over the list values using the {\Nextgeq} primitive,
and advancing to the element
coming next the one currently ``pointed to''.
At each step of the inner loop,
the heap selects the iterator that currently ``points to''
the minimum docid.
If such docid is smaller than \var{x},
then we can advance the iterator
to the successor of \var{x} by calling {\Nextgeq}
and re-heapify the heap (line 13).
Otherwise we have that such docid is
larger-than or equal-to \var{x}.
If it is equal to \var{x}, then a result is found.
Then in any case we can break the loop because either
a result was found,
or the docid is strictly larger then \var{x},
thus also every other element in the heap is larger than \var{x}.
Fig.~\ref{fig:conjunctive_search_heap_example}
details a step-by-step
example showing the behavior of the algorithm.

\begin{figure}[t]
\removelatexerror
\centering
\scalebox{\mytablescale}{\input{conjunctive_search_heap.tex}}
\mycaption{\emph{Heap-based} {\conj} algorithm.
\label{alg:conjunctive_search_heap}}
\end{figure}

\begin{figure}[t]
\includegraphics[scale=0.47]{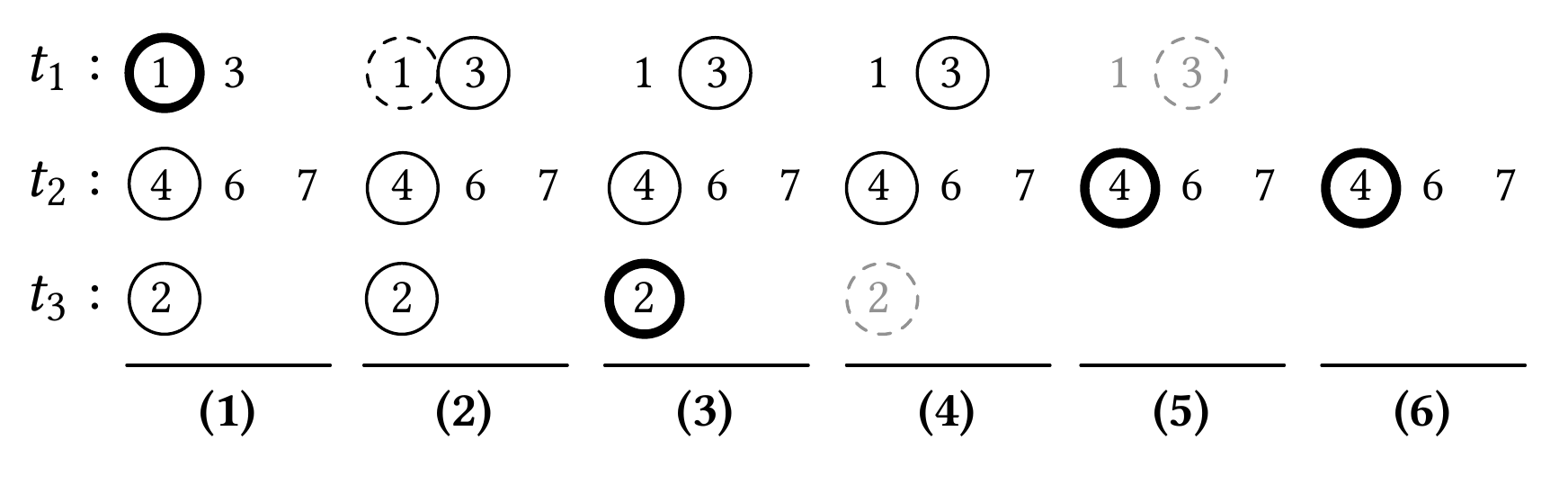}
\mycaption{The steps performed by the algorithm in
Fig.~\ref{alg:conjunctive_search_heap} for the query
\textsf{``bmw i3 s''}.
We check the elements in the
intersection between the inverted lists of \textsf{``bmw''} and \textsf{``i3''},
that are $[1,2,4]$, over the lists for the terms
\textsf{``sedan''} ($t_1:1,3$), \textsf{``sport''} ($t_2:4,6,7$) and \textsf{``sportback''} ($t_3:2$).
The elements pointed to by the iterators in the heap
are the ones circled with solid lines.
At the beginning we are checking docid 1 and, since heap's top element is 1,
that is the first result. At the second step we are now checking docid 2.
Since the heap still returns 1, we advance the iterator to $\Nextgeq_{t_1}(2) = 3$.
At step 3, the heap returns the element 2 that is another found result.
At step 4 we are now checking docid 4, thus we advance the iterator 
by calling $\Nextgeq_{t_3}(4)$. Since the inverted list of term $t_3$ has no
element larger-than or equal to $4$, then $\Nextgeq_{t_3}(4)$ will be equal to the
sentinel $\infty$
and the iterator over such list is popped-out from the heap.
The same happens for the iterator of the list $t_1$
at step 5.
The algorithm finally finds the last result 4, at step 6.
\label{fig:conjunctive_search_heap_example}}
\end{figure}

Let $m = r - \ell + 1$ be the size of the range $[\ell,r]$.
The filling and making of the heap (lines 4-6) takes $O(m)$ time.
For every element of the intersection,
we execute the inner loop (lines 9-18) that
has a worst-case
complexity of $t_{\emph{\text{check}}} = O(m\log m \times t_{\Nextgeq})$.
Therefore, the overall complexity is
$O(m + t_{\emph{\text{intersection}}} + |\var{intersection}| \times t_{\emph{\text{check}}})$.
We point out that this theoretical complexity is, however,
excessively pessimistic because $t_{\emph{\text{check}}}$ may be
very distant from its worst-case scenario, and indeed 
be even $O(1)$ when the body of the \func{else} branch 
at line 15 is executed (lines 16-18).
Also, the heap cost of $O(m \log m)$ progressively vanishes
as iterators are popped out from the data structure (line 14).
In fact, as we will better show in Section~\ref{sec:experiments},
the algorithm is pretty fast unless $m$ is very large.
Handling large values of $m$ efficiently is indeed the
problem we address in the following.

Lastly, we point out that the approach
by~\citet*{bast2006type} can be implemented on top of
this algorithm.
The crucial difference is that their algorithm makes use of
a \emph{blocked} inverted index,
with inverted lists grouped into blocks
and merged.
We will compare against their approach in Section~\ref{sec:experiments}.

\parag{Forward Search}
The approach coded in Fig.~\ref{alg:conjunctive_search_heap}
is clearly more convenient than explicitly computing
the union of all the inverted lists in $[\ell,r]$
and then searching it for every single docid belonging to the intersection.
However,
it is inefficient when the range $[\ell,r]$ is very large.
We remark that this case is actually possible
and \emph{very frequent} indeed,
because it represents the case where the user
has typed just few characters of the suffix and, potentially,
a large number of strings are prefixed by such characters.
We now discuss how to solve this problem efficiently.

The idea, illustrated in Fig.~\ref{alg:conjunctive_search_forward},
is to avoid accessing the inverted lists of the terms
in the range $[\ell,r]$
(and, thus, avoid using a heap data structure as well)
but rather check whether the terms of a completion
intersects the ones in $[\ell,r]$.
More precisely, for every completion in the intersection
we check if there is at least one term $t$ of the completion such that $t \geq \ell$ and $t \leq r$.
Given that completions do not contain
many terms (see also Table~\ref{tab:datasets}),
a simple scan of the completion suffices to implement the check
as fast as possible.
While this is not constant-time from a theoretical
point of view, in practice it is.
This idea of falling back to a forward search
was introduced by~\citet*{ji2009efficient}.

Take again the example query \textsf{``bmw i3 s''}.
We check whether the completions of docid 1, 2, and 4,
seen as integer sets, intersect the range $[7,9]$.
By looking at Table~\ref{tab:completions}, it is easy to see that
the last term id of such completions is always in $[7,9]$.

The complexity of the algorithm is then essentially dependent
from the size of the intersection and the time needed to {\extract}
a completion, that is $O(t_{\emph{\text{intersection}}} + |\var{intersection}| \times t_{\extract})$.
Compared to the heap-based algorithm in Fig.~\ref{alg:conjunctive_search_heap},
we are improving the time for checking a given docid
(and saving a factor of $O(m)$),
by relying of the efficiency of the {\extract} operation.
We clearly expect $t_{\extract}$ to be more efficient than the
worst-case complexity of $t_{\emph{\text{check}}}$ that is
$O(m\log m \times t_{\Nextgeq})$, especially for large values of $m$.
Note that although the worst-case theoretical complexity
is independent from $m$, in practice the size of $m$ influences
the probability that the test in line 6 succeeds: the larger is $m$,
the higher the probability and the faster the running time of the algorithm.

However, the behavior of the heap-based algorithm for \emph{small}
values of $m$ is not intuitive and its running time could not
necessarily be worse than having to issue many {\extract} operations
(when, for example, the test in line 11 succeeds frequently).
Again, the experimental analysis in Section~\ref{sec:experiments}
will compare the two different approaches.
Instead, it should be intuitive why this algorithm produces the same results as
the heap-based one:
they are just the ``inverted version'' of each other, i.e.,
one is using an inverted index whereas the other is using a ``forward'' approach.
Therefore, correctness is immediate.

To {\extract} a completion given its id (line 6),
we can either use a forward index or {\compfc} compression, as we
have discussed in Section~\ref{sec:data_structures}.
Using the latter method means to actually
\emph{decode} a completion, a process involving scanning
and memory-copy
operations, whereas the former technique provides immediate access to
the completion, that is $t_{\extract} = O(1)$,
at the expense of storing an additional data structure
(the forward index).
Therefore, we have a potential space/time trade-off here, that we
investigate in Section~\ref{sec:experiments}.

\begin{figure}[t]
\removelatexerror
\centering
\scalebox{\mytablescale}{\input{conjunctive_search_forward.tex}}
\mycaption{\emph{Forward} {\conj} algorithm.
\label{alg:conjunctive_search_forward}}
\end{figure}

\vspace{1cm}
\parag{Single-Term Queries}
Now we highlight another efficiency issue: the case for
single-term queries. 
We recall and remark
that such queries are always executed
when users are typing,
hence they are the most frequent case.
This motivates the need for having a specific algorithm for
their resolution.

Single-term queries represent a special case in that
the prefix is \emph{empty} (we only have the suffix).
This means that there is no intersection over which to
iterate, rather every single docid from 1 to $|\seq|$
would have to be considered by both algorithms coded in Fig.~\ref{alg:conjunctive_search_heap}
and~\ref{alg:conjunctive_search_forward}.
This makes them very inefficient on such queries.
In this case, the ``classic'' approach of finding
the $k$ smallest elements from the
inverted lists in the range $[\ell,r]$
with a heap is more efficient than checking every docid
(using a similar approach to that coded in Fig.~\ref{alg:conjunctive_search_heap}).
However, it is still slow on large ranges
because an iterator for every inverted list
in the range $[\ell,r]$ has to be instantiated
and pushed onto the heap.

We can elegantly solve this problem
with another {\rmq} data structure.
Let us consider the list \var{minimal}, where
$\var{minimal}[i]$ is the
first docid of the $i$-th inverted list.
(In other words, \var{minimal} is the ``first column'' of the
inverted index.)
If we build a {\rmq} data structure on such list,
$\rmq(\ell,r)$ identifies the 
inverted list from which the minimum docid is returned.
Therefore, we instantiate an iterator on such list and
push onto the heap its \emph{next} docid
along with the left and right sub-ranges.
We proceed recursively as explained in the previous section.
(Now, if the element at the top of the heap comes
from an iterator we do not push left and right sub-ranges.)
The key difference with respect to the ``classic'' heap-based
algorithm mentioned above,
is that an iterator
is instantiated over an inverted list \emph{if and only if}
an element has to be returned from it.

For the example in Table~\ref{tab:inverted_index},
the \var{minimal} list will be $[$6, 3, 1, 1, 7, 3, 1, 4, 2, 5$]$
and, if the single-term query is \textsf{``s''},
then we ask for {\rmq} over $\var{minimal}[7,9]$.
Assume $k=3$.
The first returned docid is therefore 1, the first
for the inverted list of the term ``\textsf{sedan}''.
We pushed onto the heap the next id from such list, 3,
as well as the right sub-range $[8,9]$.
The element at the top of the heap is now 2, the first
for the inverted list of the term ``\textsf{sportback}''.
There are no more docids from such list, thus we remove
the sub-range $[8,9]$ and add the sub-range $[8,8]$.
We finally return the id 3, again from the list
of the term ``\textsf{sedan}''.
Observe that the iterator on the inverted list of the
term id 8 (``\textsf{sport}'', in this case)
is never instantiated.

\newpage
 
\section{Experiments}\label{sec:experiments}

In this section we report on the experiments
we conducted to assess the efficiency and the effectiveness
of the described QAC algorithms.
The experiments are organized as follows.
We first benchmark and tune the data structures used by the algorithms
in Section~\ref{sec:exp_data_structures}.
With the tuning done, we then compare the efficiency of various options
to perform {\conj},
also with respect to the efficiency of {\pref}, in Section~\ref{sec:efficiency}.
We then discuss effectiveness and memory footprint of the various
solutions in Section~\ref{sec:effectiveness} and~\ref{sec:space}
respectively.

\begin{table}[t]
\centering
\mycaption{Dataset statistics.}
\scalebox{\mytablescale}{\begin{tabular}{l rrr}
\toprule
Statistic & {\aol} & {\msn} & {\ebay} \\
\midrule
Queries & \num{10142395} & \num{7083363} & \num{7295104} \\
Uncompressed size in MiB & 299 & 208 & 189 \\
Unique query terms & \num{3825848} & \num{2590937} & \num{323180} \\
Avg. num. of chars per term & 14.58 & 14.18 & 7.32 \\
Avg. num. of queries per term & 7.87 & 8.15 & 73.02 \\
Avg. num. of terms per query & 2.99 & 2.99 & 3.24 \\
\bottomrule
\end{tabular}
}
\label{tab:datasets}
\end{table}

\parag{Datasets}
We used three large real-world query logs in English:
{\aol}~\cite{aol} and {\msn}~\cite{msn}
(both available at
\url{https://jeffhuang.com/search_query_logs.html}),
and {\ebay} that is a proprietary collection of
queries collected during the year 2019 from the US .com site.
We do not apply any text processing to the logs,
such as capitalization, but index the strings as given
in order to ensure accurate reproducibility of our results.
For {\aol} and {\msn}, the score of a query is the number
of times the query appears in the log
(i.e., its frequency count)~\cite{hsu2013space};
for {\ebay}, the score is assigned by some machine learning facility
that is irrelevant for the scope of this paper.
As already mentioned, integer ids (docids)
have been assigned to queries
in decreasing score order. Ties are broken lexicographically.
Table~\ref{tab:datasets} summarizes the statistics.

\parag{Experimental Setting}
Experiments were performed on a server machine equipped
with Intel i9-9900K cores (@3.60 GHz),
64 GB of RAM DDR3 (@2.66 GHz) and running Linux 5 (64 bits).

For researchers interested in replicating the results
on public datasets, we provide the C++ implementation at
\url{https://github.com/jermp/autocomplete}.
We used that implementation to obtain the results
discussed in the paper.
The code was compiled with \textsf{gcc} 9.2.1
with all optimizations enabled,
that is with flags \texttt{-O3} and \texttt{-march=native}.

The data structures were flushed to disk after construction and loaded in
memory to be queried.
The reported timings are average values among 5 runs of the same experiment.
All experiments run on a single CPU core.
We use $k=10$ for all experiments.

\subsection{Tuning the Data Structures}\label{sec:exp_data_structures}

For the experiments in this section, we used the (larger)
{\aol} dataset given that consistent results were also
obtained for {\msn} and {\ebay}.

\parag{The Dictionary}
As explained in Section~\ref{sec:data_structures},
we represent the dictionary using a 2-level data structure
compressed with Front Coding ({\compfc}).
We are interested in benchmarking the time for
three operations, namely
{\extract}, {\locate}, and {\locateprefix},
by varying the bucket size that directly controls
the achievable space/time trade-off.
The result of the benchmark is reported in Table~\ref{tab:aol.fc_dictionary}.
The timings, expressed in $\mu$sec per string,
are recorded by executing 100,000 queries and
computing the average.
Such queries are strings belonging to the dictionary and
shuffled at random to avoid locality of access.
To benchmark the operation {\locateprefix}, we retain
0\%, 25\%, 50\% and 75\% of the characters of a given input string
(the case for 100\% would correspond to a {\locate} operation;
in the case of 0\%, we always retain 1 single character instead of 0).
As we can see from the results reported in the table,
the space decreases but the time increases
for increasing values of bucket size.
The {\extract} operation is roughly $4\times$ faster than {\locate}.
The timings for {\locateprefix} are pretty much the same for
all percentages except 0\%: in that case strings comparisons are much
faster, resulting in a better execution time.
For all the following experiments, we choose a bucket size of 16
that yields a good space/time trade-off:
{\extract} takes 0.1 $\mu$sec on average, with {\locate} and
{\locateprefix} around half of a microsecond and,
compared to the size of the uncompressed file which is 56.85 MiB,
{\compfc} offers a compression ratio of approximately $1.69\times$.
(On {\msn} and {\ebay}, the compression ratio are
$1.67\times$ and $1.61\times$ respectively.)

\begin{table}[t]
\centering
\mycaption{Front-coded dictionary benchmark on the {\aol} dataset,
by varying bucket size.
Timings are in $\mu$sec per string.
The size of the uncompressed file is 56.85 MiB,
that is an average of 15.58 bytes per string (bps).}
\scalebox{\mytablescale}{\begin{tabular}{ccrcccccc}
\toprule
\multirow{2}{*}{Bucket} & \multirow{2}{*}{MiB} & \multirow{2}{*}{bps} & \multirow{2}{*}{{\extract}} & \multirow{2}{*}{{\locate}} & \multicolumn{4}{c}{{\locateprefix}} \\
\cmidrule(lr){6-9}
size & & & & & 0\% & 25\% & 50\% & 75\% \\
\midrule
4 & 40.95 & 11.22 & 0.12 & 0.46 & 0.15 & 0.67 & 0.76 & 0.66 \\
8 & 36.35 & 9.96 & 0.11 & 0.43 & 0.17 & 0.62 & 0.65 & 0.58 \\
16 & 33.64 & 9.22 & 0.10 & 0.41 & 0.18 & 0.61 & 0.62 & 0.57 \\
32 & 32.16 & 8.81 & 0.12 & 0.44 & 0.22 & 0.69 & 0.69 & 0.65 \\
64 & 31.39 & 8.60 & 0.16 & 0.54 & 0.57 & 0.89 & 0.92 & 0.89 \\
128 & 30.99 & 8.49 & 0.24 & 0.74 & 0.51 & 1.21 & 1.31 & 1.30 \\
256 & 30.79 & 8.44 & 0.42 & 1.20 & 0.96 & 2.07 & 2.23 & 2.24 \\
\bottomrule
\end{tabular}
}
\label{tab:aol.fc_dictionary}
\end{table}

\begin{figure}[t]
\centering
\subfloat[{\locateprefix}] {
\includegraphics[scale=0.72]{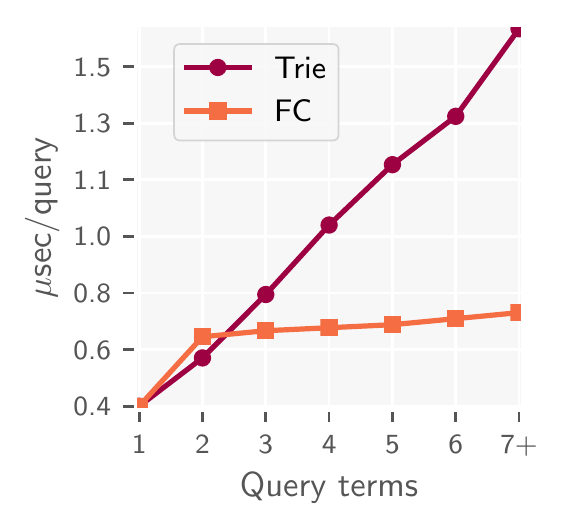}
\label{fig:aol.locate_prefix}
}
\subfloat[{\rmq}] {
\includegraphics[scale=0.72]{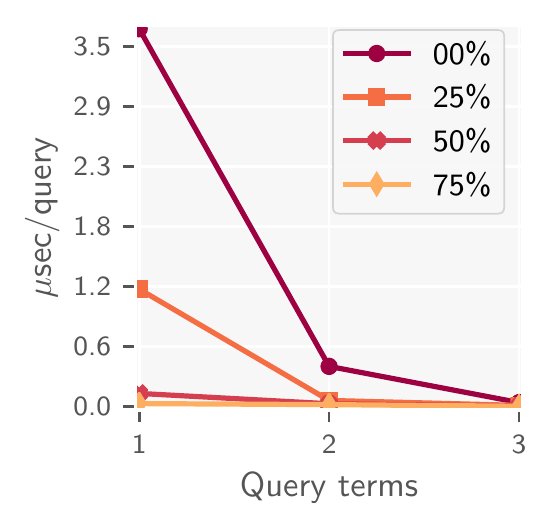}
\label{fig:aol.rmq}
}
\mycaption{Timings for (a) {\locateprefix} and (b) {\rmq}
in $\mu$sec per string on {\aol}.
Results for {\compfc} are relative
to a bucket size of 16 strings.}
\end{figure}


\parag{The Completions}
We now compare the two distinct approaches of representing the completions
with a trie or Front Coding.
Regarding the trie, we recall from Section~\ref{sec:data_structures}
that it is represented by four sorted integer sequences.
We follow the design recommended in~\cite{pibiri2019handling,pibiri2017efficient}
of using Elias-Fano to represent nodes and pointers for its fast,
namely constant-time, random access algorithm and powerful search capabilities.
For the same reasons, we also adopt Elias-Fano to compress the left extremes
and range sizes.
With Elias-Fano compression, the trie
takes a total of 88.80 MiB that is 9.18 bytes
per completions (bpc).
Most of the space is spent, not surprisingly, in the encoding
of the nodes: 6.57 bpc (71.6\%).
Pointers take 0.84 bpc (9.17\%), left extremes take 1.08 bpc (11.73\%) and
range sizes take 0.69 bpc (7.5\%).
The completions compressed with {\compfc}, using a bucket size of 16,
take 97.98 MiB, i.e., 10.13 bpc.
Thus the trie takes 9.4\% less space than {\compfc}.


To record the time for the {\locateprefix} operation,
we partitioned the completions by the number of terms $d$,
for $d$ from 1 to 6. All completions having $d \geq 7$ terms (7+)
are placed in the same partition.
From each partition, we then sample 100,000
queries at random.
We firs observed that the time is
pretty much independent from the size of the suffix
because the average number of characters per term is very low.
(Basically, 14 for both {\aol} and {\msn}, and 7 for {\ebay}.
See Table~\ref{tab:datasets}.)
Therefore, the only influence comes from the number of
query terms and we show the result in Fig.~\ref{fig:aol.locate_prefix}.

While the {\comptrie} query time constantly increases by $\approx$200 nanoseconds
per level (basically, 2 cache misses per level),
the query time for {\compfc} is almost insensitive to the size of the query.
Therefore as expected, the {\comptrie} is beaten by
{\compfc} as query length increases due to cache-misses.
As a net result, better cache efficiency
paired with fast decoding
makes {\compfc} roughly $2\times$ faster than the {\comptrie}
for queries having more than 4 terms.

\parag{Range-Minimum Queries}
The timings for {\rmq} are reported in Fig.~\ref{fig:aol.rmq}.
As it is intuitive, the timing strongly depends on the size of the
range. Such size is exponentially decreasing when
both the number of terms and the percentage of characters
retained from the suffix increases.
As a matter of fact, the {\rmq} time is practically negligible
from 3 terms onwards.


\begin{table}[t]
\centering
\mycaption{Inverted index compression benchmark on {\aol}
in average bits per integer (bpi).
}
\scalebox{\mytablescale}{\begin{tabular}{c ccccccc}
\toprule

Method & {\bic} & {\dint} & {\pef} & {\ef} & {\optvb} & {\vb} & {\simples} \\
\midrule
bpi & 14.14 & 15.08 & 15.10 & 17.15 & 17.33 & 20.95 & 21.74 \\



\bottomrule
\end{tabular}
}
\label{tab:aol.ii_compression}
\end{table}

\parag{Inverted Index Compression}
For the QAC problem,
the inverted lists are very short on average because the completions
themselves comprise only few terms (see Table~\ref{tab:datasets}).
Therefore, we cannot expect a great deal of compression
effectiveness as, for example, the one for Web pages~\cite{survey}.
Nonetheless, we experimented with several compressors, such as:
Binary Interpolative Coding ({\bic})~\cite{2000:moffat.stuiver},
dictionary-based encoding ({\dint})~\cite{pibiri2019fast},
Elias-Fano ({\ef})~\cite{Fano71,Elias74},
partitioned Elias-Fano ({\pef})~\cite{ottaviano2014partitioned},
Variable-Byte paired with SIMD instructions~\cite{2015:plaisance.kurz.ea},
optimally-partitioned Variable-Byte ({\optvb})~\cite{pibiri2019optimally},
and {\simples}~\cite{zlt08www}.
A description of all such compression
methods can be found in the recent survey on the topic~\cite{survey}.
We report the average number of bits spent per represented integer (bpi)
by such methods in
Table~\ref{tab:aol.ii_compression}.
We also collected the timings to compute intersections
by varying the number of query terms (using the same queries as used
for the {\locateprefix} experiment
in order to compute intersections among inverted lists
relative to terms that co-occur in real completions).
Apart from {\bic} that is roughly $3\times$ slower,
all other techniques offer similar efficiency. 

In conclusion, we choose Elias-Fano ({\ef})
to compress the inverted lists
for its good space effectiveness, efficient query time and
compact implementation.
We respect to the uncompressed case, {\ef} saves
roughly 50\% of the space.




\begin{table*}[t]
\centering

\mycaption{Top-10 {\conj} query timings in $\mu$sec per query, by varying
query length and percentage of the last query token.}

\subfloat[{\aol}]{
\scalebox{0.8}{\hspace{-1cm}\setlength{\tabcolsep}{3pt}
\begin{tabular}{cc rrrrrrr}

\cmidrule[0.7pt](lr){3-9}

& & \multicolumn{7}{c}{Query terms} \\ 
\cmidrule(lr){3-9}
& \% & 1 & 2 & 3 & 4 & 5 & 6 & 7+ \\ 

\cmidrule[0.7pt](lr){2-2}
\cmidrule(lr){3-9}

\multirow{4}{*}{\rotatebox[origin=c]{90}{{\one}}} 
& 0 & 4 & 5 & 22 & 30 & 24 & 24 & 16 \\ 
& 25 & 2 & 97 & 70 & 41 & 30 & 25 & 16 \\ 
& 50 & 0 & 149 & 77 & 48 & 30 & 25 & 16 \\ 
& 75 & 0 & 150 & 76 & 48 & 30 & 25 & 16 \\ 

\cmidrule[0.7pt](lr){2-2}
\cmidrule(lr){3-9}

\multirow{4}{*}{\rotatebox[origin=c]{90}{{\two}}} 
& 0 & 5 & 15 & 27 & 30 & 24 & 24 & 16 \\ 
& 25 & 3 & 251 & 110 & 45 & 31 & 25 & 16 \\ 
& 50 & 1 & 370 & 121 & 56 & 31 & 25 & 16 \\ 
& 75 & 0 & 375 & 121 & 57 & 32 & 25 & 16 \\

\cmidrule[0.7pt](lr){2-2}
\cmidrule(lr){3-9}

\multirow{4}{*}{\rotatebox[origin=c]{90}{{\three}}} 
& 0 & \small{55,537} & \small{29,189} & \small{30,498} & \small{22,431} & \small{17,713} & \small{16,474} & \small{13,312} \\ 
& 25 & 474 & 623 & 957 & 485 & 376 & 378 & 299 \\ 
& 50 & 1 & 251 & 178 & 251 & 229 & 123 & 178 \\ 
& 75 & 0 & 226 & 162 & 240 & 219 & 116 & 173 \\

\cmidrule[0.7pt](lr){2-2}
\cmidrule(lr){3-9}

\multirow{4}{*}{\rotatebox[origin=c]{90}{{\four}}} 
& 0 & 286 & \small{2,718} & \small{1,673} & 965 & 634 & 503 & 413 \\ 
& 25 & 11 & 184 & 223 & 276 & 258 & 221 & 192 \\ 
& 50 & 10 & 126 & 185 & 270 & 250 & 217 & 186 \\ 
& 75 & 6 & 116 & 178 & 268 & 248 & 216 & 184 \\ 

\cmidrule[0.7pt](lr){2-2}
\cmidrule[0.7pt](lr){3-9}

\end{tabular}}
}
\subfloat[{\msn}]{
\scalebox{0.8}{\setlength{\tabcolsep}{3pt}
\begin{tabular}{rrrrrrr}

\cmidrule[0.7pt](lr){1-7}
\multicolumn{7}{c}{Query terms} \\ 
\cmidrule(lr){1-7}
1 & 2 & 3 & 4 & 5 & 6 & 7+ \\ 
\cmidrule(lr){1-7}
4 & 5 & 14 & 15 & 11 & 10 & 7 \\ 
1 & 39 & 34 & 18 & 13 & 10 & 7 \\ 
0 & 56 & 38 & 19 & 13 & 10 & 8 \\ 
0 & 57 & 37 & 19 & 12 & 10 & 7 \\ 
\cmidrule(lr){1-7}
5 & 15 & 17 & 15 & 11 & 10 & 7 \\ 
2 & 101 & 51 & 19 & 13 & 10 & 8 \\ 
1 & 137 & 58 & 21 & 13 & 10 & 7 \\ 
0 & 137 & 57 & 21 & 13 & 10 & 7 \\ 
\cmidrule(lr){1-7}
\small{7,626} & \small{12,459} & \small{11,964} & \small{8,921} & \small{6,164} & \small{5,749} & \small{5,686} \\ 
353 & 252 & 256 & 282 & 170 & 192 & 125 \\ 
10 & 73 & 70 & 109 & 84 & 66 & 54 \\ 
1 & 61 & 62 & 83 & 80 & 63 & 51 \\ 
\cmidrule(lr){1-7}
53 & \small{1,626} & 915 & 477 & 307 & 270 & 237 \\ 
10 & 90 & 109 & 127 & 111 & 111 & 90 \\ 
7 & 53 & 97 & 122 & 107 & 108 & 87 \\ 
4 & 46 & 95 & 121 & 106 & 106 & 85 \\ 
\cmidrule[0.7pt](lr){1-7}
\end{tabular}}
}
\subfloat[{\ebay}]{
\scalebox{0.8}{\setlength{\tabcolsep}{3pt}
\begin{tabular}{rrrrrrr}
\cmidrule[0.7pt](lr){1-7}
\multicolumn{7}{c}{Query terms} \\ 
\cmidrule(lr){1-7}
1 & 2 & 3 & 4 & 5 & 6 & 7+ \\ 
\cmidrule(lr){1-7}
3 & 6 & 53 & 80 & 96 & 146 & 94 \\ 
1 & 125 & 115 & 111 & 112 & 152 & 95 \\ 
1 & 214 & 131 & 113 & 114 & 151 & 95 \\ 
1 & 239 & 132 & 114 & 113 & 150 & 95 \\ 
\cmidrule(lr){1-7}
4 & 16 & 59 & 81 & 96 & 146 & 96 \\ 
3 & 258 & 133 & 115 & 113 & 152 & 96 \\ 
2 & 444 & 153 & 117 & 115 & 151 & 94 \\ 
2 & 494 & 156 & 119 & 114 & 150 & 94 \\ 
\cmidrule(lr){1-7}
120 & \small{4,799} & \small{6,391} & \small{4,618} & \small{3,566} & \small{1,945} & 971 \\ 
43 & 854 & \small{1,392} & \small{1,28} & 904 & 727 & 331 \\ 
41 & 603 & \small{1,213} & 895 & 835 & 687 & 314 \\ 
41 & 594 & \small{1,217} & 909 & 840 & 688 & 312 \\ 
\cmidrule(lr){1-7}
15 & \small{2,909} & \small{2,827} & \small{1,756} & \small{1,371} & 821 & 417 \\ 
9 & 638 & 790 & 580 & 553 & 543 & 303 \\ 
11 & 454 & 694 & 513 & 530 & 537 & 297 \\ 
12 & 454 & 698 & 517 & 529 & 536 & 297 \\ 
\cmidrule[0.7pt](lr){1-7}
\end{tabular}}
}

\label{tab:conj.timings}
\end{table*}

\begin{table*}[t]
\centering

\mycaption{Percentage of better scored results returned by {\conj}
wrt those returned by {\pref} for top-10 queries.}

\subfloat[{\aol}]{
\scalebox{\mytablescale}{\hspace{-0.5cm}\begin{tabular}{c rrrrrrr}

\cmidrule[0.7pt](lr){2-8}

& \multicolumn{7}{c}{Query terms} \\ 
\cmidrule(lr){2-8}
\% & 1 & 2 & 3 & 4 & 5 & 6 & 7+ \\ 

\cmidrule[0.7pt](lr){1-1}
\cmidrule(lr){2-8}

0  & 17 & 107 & 207 & 327 & 295 & 270 & 270 \\ 
25 & 19 & 178 & 246 & 373 & 298 & 155 & 356 \\ 
50 & 23 & 227 & 302 & 440 & 364 & 213 & 524 \\ 
75 & 41 & 282 & 362 & 504 & 424 & 257 & 882 \\ 

\cmidrule[0.7pt](lr){1-1}
\cmidrule[0.7pt](lr){2-8}

\end{tabular}}
}
\subfloat[{\msn}]{
\scalebox{\mytablescale}{\begin{tabular}{rrrrrrr}

\cmidrule[0.7pt](lr){1-7}

\multicolumn{7}{c}{Query terms} \\ 
\cmidrule(lr){1-7}
1 & 2 & 3 & 4 & 5 & 6 & 7+ \\ 

\cmidrule(lr){1-7}

27 & 139 & 252 & 283 & 325 & 206 & 248 \\ 
23 & 231 & 310 & 297 & 333 & 190 & 200 \\ 
27 & 243 & 313 & 320 & 359 & 251 & 208 \\ 
44 & 284 & 364 & 357 & 407 & 319 & 236 \\ 

\cmidrule[0.7pt](lr){1-7}

\end{tabular}}
}
\subfloat[{\ebay}]{
\scalebox{\mytablescale}{\begin{tabular}{rrrrrrr}

\cmidrule[0.7pt](lr){1-7}

\multicolumn{7}{c}{Query terms} \\ 
\cmidrule(lr){1-7}
1 & 2 & 3 & 4 & 5 & 6 & 7+ \\ 

\cmidrule(lr){1-7}

48 & 85 & 102 & 130 & 136 & 167 & 159 \\ 
55 & 89 & 103 & 133 & 146 & 152 & 129 \\ 
50 & 86 & 104 & 133 & 148 & 152 & 138 \\ 
50 & 87 & 106 & 132 & 149 & 153 & 136 \\ 

\cmidrule[0.7pt](lr){1-7}

\end{tabular}}
}

\label{tab:effectiveness}
\end{table*}

\subsection{Efficiency}\label{sec:efficiency}

With the tuning of the data structures done,
we are now ready to discuss the efficiency of the main
building blocks that we may use to implement a QAC algorithm,
namely {\pref} and {\conj}, as well as
that of the (minor) steps of
parsing the query and reporting the strings.

In all the subsequent experiments,
we are going to use the following methodology to measure
the query time of the indexes.
For both {\aol} and {\msn},
we sampled 1,000 queries at random from each set of completions
having $d = 1,\ldots,6$ and $d \geq 7$ terms (7+),
and use these completions as queries.
We built the indexes by \emph{excluding} such queries.
For {\ebay}, we took a log of 2.7 million queries
collected in early 2020, again from the US .com site,
and sampled 7,000 queries as explained above.
The queries are answered in random order
(i.e., in no particular order)
to avoid locality of access.

\parag{Conjunctive-search}
We compare the following algorithms for {\conj}:
the \emph{heap-based} (Fig.~\ref{alg:conjunctive_search_heap})
and indicated as {\three},
the two implementations of the \emph{forward-based}
(Fig.~\ref{alg:conjunctive_search_forward})
that respectively use a forward index ({\one}) and Front Coding ({\compfc}),
and the {\four} index by~\citet*{bast2006type}\footnote{The {\four} index
depends on a parameter $c$ that controls the degree of associativity
of the inverted lists.
This parameter affects the trade-off
between space and time~\cite{bast2006type}.
We built indexes for different values of $c$,
and found that
the value $c = 10^{-4}$ gives the best
space/time trade-off. Therefore, this is the value of $c$
we used for the following experiments.}.
The comparison is reported in Table~\ref{tab:conj.timings}.
The first thing to note is that the impact of the different
solutions is very consistent
across the datasets (although the timings are different),
therefore all considerations expressed
in the following apply to all datasets.

\begin{itemize}
\item As foreseen in Section~\ref{sec:conjunctive},
{\three} is several order of magnitude slower than all other
approaches whenever the lexicographic range of the suffix is
very large as it happens for the 0\% row.
Although this latency may not be
acceptable for real-time performance,
observe the sharp drop in the running time as soon as we
have longer suffixes ($\geq 25\%$): we pass from milliseconds to
a few hundred microseconds.
{\four} protects against the worst-case behaviour of {\three},
thus confirming the analysis in the original paper~\cite{bast2006type}.
However, since {\three} is faster than {\four} at performing
list intersection, it is indeed competitive with {\four} for sufficiently long suffixes
(e.g., $\geq 50\%$).

\item The solutions {\one} and {\two} significantly
outperform {\three} and {\four} by a wide margin for
the reasons we explained in Section~\ref{sec:conjunctive}.
There is not a marked difference between {\one} and {\two},
except for the case with two query terms.
This is the case
where the prefix comprises only one term, thus every docid
in its inverted list must be checked until $k$ results are
found or the list is exhausted.
{\one} is faster
than {\two} in this case because the many {\extract} operations
performed over the strings compressed with Front Coding impose an overhead,
resulting in a slowdown
with respect to {\one} (and, sometimes, {\three} as well).
Interestingly enough, this
slowdown progressively vanishes as fewer results
need to be checked, such as with 3 or more query terms.
(This also suggests that, when using {\two}, we could 
switch to {\three} for sufficiently long suffixes
and two query terms.)
The case with two query terms also sheds light on the influence
of the suffix size for {\one} and {\two}.
Although the worst-case complexity is independent from it because
\emph{all} docids are checked in the worst-case,
in practice the running time
increases with the suffix size because the test
performed in line 6 of the algorithm in Fig.~\ref{alg:conjunctive_search_forward}
becomes progressively more selective.
In fact, working with a small
lexicographic range lowers the probability that a completion
has at least one term in the range.

\item Lastly, consider the case for 1 query term.
The solutions using RMQ on the minimal docids, {\one} and {\two},
keep the response time orders of magnitude lower compared to {\three} and {\four}
when the suffix is very short (0\% -- 25\%).
Again, observe the drop in the running time as soon
longer suffixed are specified ($\geq 50\%$).
This is especially true for {\three} and {\four}
because only few inverted lists are accessed.
\end{itemize}

\parag{Prefix-search}
As we discussed in Section~\ref{sec:query_processing},
{\pref} comprises two {\locateprefix} operations:
one performed on the dictionary data structure that,
for a choice
of bucket size equal to 16, costs 0.2 -- 0.6 $\mu$sec
per string (Table~\ref{tab:aol.fc_dictionary});
the other performed
on the set of completions,
for a cost of 0.4 -- 1.7 $\mu$sec per string
if we use a trie,
or 0.4 -- 0.7 $\mu$sec per string
if we use Front Coding (Fig.~\ref{fig:aol.locate_prefix}).
Therefore, summing together these contributions,
we have that {\pref} is supported in either:
0.6 -- 2.4 $\mu$sec per query;
or even less if we allow more space, i.e.,
in 0.6 -- 1.4 $\mu$sec per query
for 9.4\% of space more.
Lastly, to this cost we have also to add that
of {\rmq} that, as seen in Fig.~\ref{fig:aol.rmq},
is relevant only for queries having 1 and 2 terms
with a few characters typed at the end.

The timings for {\conj}, as reported
in Table~\ref{tab:conj.timings}, are far from
being competitive from those of {\pref},
being actually orders of magnitude larger especially
on shorter queries.
This is not surprising given that {\conj}
involves querying an inverted index and accessing
other data structures, like a forward index (for {\one})
or a compressed set of strings (for {\two}).
The use of {\conj} is, however,
motivated by its increased effectiveness
as we are going to discuss next.

\parag{Other Costs}
Further costs include that of parsing the query (i.e.,
looking-up each term in the dictionary) and reporting
the actual strings given a list of top-$k$ docids.
Both operations add a small cost -- always
below 2 $\mu$sec per query, even in the case of very long
queries and many reported results.

\subsection{Effectiveness}\label{sec:effectiveness}

We now turn our attention to the comparison 
between {\pref} and {\conj}
by considering
their respective effectiveness.
As already pointed out, {\pref} is cheaper from
a computational point of view but has limited
discovery power, i.e., its matches are restricted
to string that are prefixed by the user input.
A simple and popular metric to asses the effectiveness
of different QAC algorithms, is \emph{coverage}~\cite{cao2008context,jones2006generating},
defined to be the fraction of queries for which the algorithm
returns at least one result.
However, coverage alone is little informative~\cite{bhatia2011query}
because it is not able to capture the \emph{quality}
of the returned results.
In the example
of Fig.~\ref{fig:screenshots}b-c,
both {\pref} and {\conj} return 10 results
but 8 of those returned by {\conj} have a better score
than those returned by {\pref}.
Therefore, we use a different metric.

We consider the set of completions' scores
for a query $q$ as given by both {\conj} and {\pref},
say $\mathcal{S}_c(q)$ and $\mathcal{S}_p(q)$ respectively.
Clearly, {\conj} returns at least the same
number of results as {\pref},
that is $|\mathcal{S}_c(q)| \geq |\mathcal{S}_p(q)|$.
Effectiveness is measured in the number of results 
returned by {\conj} that have
a better score than those returned by {\pref}.
Since for every element in $\mathcal{S}_p(q)$
we can always find an element in $\mathcal{S}_c(q)$
that has the same or a better score,
the effectiveness value for the query $q$
is $|\mathcal{S}_c(q) \setminus \mathcal{S}_p(q)|$.
We say that {\conj} returns
$|\mathcal{S}_c(q) \setminus \mathcal{S}_p(q)| / |\mathcal{S}_p(q)| \times 100\%$
better results
than {\pref} for query $q$.
For example, if the sets of scores are
$\mathcal{S}_p(q) = \{$182, 203, 344, 345$\}$
and $\mathcal{S}_c(q) = \{$123, 182, 198, 203, 344, 345$\}$,
then {\conj} found 2 more matches with better score
than those returned by {\pref},
those having score 123 and 198, hence 50\% better results.

In Table~\ref{tab:effectiveness} we report the percentage
of better results
over the same query logs used to generate Table~\ref{tab:conj.timings}.
The numbers confirm that {\conj} is a lot
more effective than {\pref} because the percentage
of better results is always well above 80\% for queries
involving more than one term.
For example,
over the {\ebay} dataset for queries having 2 terms
and by retaining 50\% of the last token,
{\conj} found 4,062 results \emph{more} than
the 4,711 found by {\pref} (for a total of 8,773 results), i.e., 86.2\% more results.

For single-term queries the possible completions for {\pref}
are many -- especially for small suffixes -- thus the difference
with respect to {\conj} is less marked.

\begin{table}[t]
\centering
\mycaption{Space usage in total MiB and bytes per completion (bpc).}
\scalebox{\mytablescale}{\begin{tabular}{l cc cc cc}
\toprule
 & \multicolumn{2}{c}{{\aol}} & \multicolumn{2}{c}{{\msn}} & \multicolumn{2}{c}{{\ebay}} \\
\cmidrule(lr){2-3}
\cmidrule(lr){4-5}
\cmidrule(lr){6-7}
 & MiB & bpc & MiB & bpc & MiB & bpc \\ 
\midrule
{\one}   & 312 & 32.28 & 218 & 32.32 & 168 & 24.14 \\
{\two}   & 266 & 27.51 & 185 & 27.42 & 140 & 20.13 \\
{\three} & 254 & 26.25 & 177 & 26.25 & 139 & 19.99 \\
{\four}  & 275 & 28.48 & 191 & 28.26 & 157 & 22.50 \\
\bottomrule
\end{tabular}
}
\label{tab:space}
\end{table}

\subsection{Space Usage}\label{sec:space}

We now discuss the space usage of the various solutions,
summarized in Table~\ref{tab:space} as total MiB
and bytes per completion (bpc).

The solution taking less space is {\three} and the one taking
more is {\one}: the difference between these two is
19\% on both {\aol} and {\msn}; 17\% on {\ebay}.
The space effectiveness of the other two solutions,
{\two} and {\four}, stand in between that
of {\three} and {\one}.

Now, starting from the space breakdown of {\one},
we discuss some details.
The dictionary component takes 10 -- 11\% of the total space
of {\aol} and {\msn}, but only 1\% for {\ebay}.
This is not surprising given that {\ebay} has (more than)
$10\times$ less unique query terms than {\aol}
(see also Table~\ref{tab:datasets}).
The completions take a significant fraction of the total space,
i.e., 28 -- 29\%; the RMQ data structure takes just 13 -- 14\%.
The inverted and forward index components are expensive,
requiring 20 -- 22\% and 27 -- 34\% respectively.
The {\two} solution takes less space than {\one} --
15\% less space on average --
because
it eliminates the forward index (although it uses Front Coding
to represent the completions that is slightly less effective
than the trie data structure).
Then, {\three} takes even less space than {\two} because it does
not build an additional RMQ data structure on the minimal docids.
Lastly, {\four} introduces some redundancy in the representation
of the inverted index component, as term ids are
needed to differentiate the elements of unions of inverted lists.

In conclusion, taking a look back at the uncompressed size
reported in Table~\ref{tab:datasets}, we can say that the
presented techniques allow efficient and effective search
capabilities with (approximately) \emph{the same or even less space}
as that of the original collections.

\section{Conclusions}\label{sec:conclusions}

In this work we explored the efficiency/effectiveness
spectrum of a \emph{multi-term prefix-search} query mode
-- referred to as the \emph{{\conj}} query mode.
The algorithm empowers the new 
implementation of eBay's query auto-completion system.
From the experimental evaluation presented in this
work on publicly available datasets, like {\aol} and {\msn},
and from our experience with eBay's data,
we can formulate the following conclusions.

\begin{itemize}
\item Conjunctive-search overcomes the limited effectiveness of
{\pref} by returning more and
better scored results.
\item While {\pref} is very fast,
requiring less then 3 $\mu$sec per query on average,
{\conj} is more expensive and costs between 4 and 500 $\mu$sec per query
depending on the size of the query.
However, we find this convenient at eBay
given its (much) increased effectiveness.
We adopt several optimization for {\conj},
including the use of a forward index ({\one}),
Front Coding ({\two}) compression, and RMQ.
\item The solution {\one} takes on average 15\% more space
than {\two} but it is faster on shorter queries (2, 3 terms).
\item Both {\one} and {\two} substantially outperform
the use of a classical as well as blocked inverted index
with small extra or even less space.
\item It is lastly advised to build RMQ succinct data structures
to lower the query times in case of single-term queries.
\end{itemize}


\begin{acks}
This work was partially supported by the BIGDATAGRAPES (EU H2020 RIA, grant agreement N\textsuperscript{\b{o}}780751), the ``Algorithms, Data Structures and Combinatorics for Machine Learning'' (MIUR-PRIN 2017), and the OK-INSAID (MIUR-PON 2018, grant agreement N\textsuperscript{\b{o}}ARS01\_00917) projects.
\end{acks}

\renewcommand{\bibsep}{3.0pt}
\balance
\bibliographystyle{ACM-Reference-Format}
\bibliography{bibliography}

\end{document}